\documentclass[journal]{new-aiaa}
\usepackage[utf8]{inputenc}
\usepackage{xcolor}
\usepackage{graphicx}
\usepackage{amsmath,comment}
\usepackage{longtable,tabularx}
\title{Low-Reynolds-number aerodynamic characteristics of airfoils with piezocomposite trailing control surfaces}

\author{Kai Zhang\thanks{Postdoctoral research associate. Department of Mechanical and Aerospace Engineering. Email: kai.zhang3@rutgers.edu.}, Bharg Shah \thanks{Graduate student. Department of Mechanical and Aerospace Engineering. Email: bjs298@soe.rutgers.edu.} and Onur Bilgen \thanks{Associate professor. Department of Mechanical and Aerospace Engineering. Email: o.bilgen@rutgers.edu.}}
\affil{Rutgers University, Piscataway, 08854, NJ}

\begin{document}

\maketitle

\begin{abstract}
Morphing wings comprised of fixed leading sections with piezocomposite trailing control surfaces have emerged as a novel active control technique for unmanned aerial vehicles.
However, the wake dynamics and aerodynamic performance of such hybrid airfoil configuration has not been thoroughly investigated.
In this paper, direct numerical simulations of two-dimensional flows over hybrid airfoils comprised of NACA 0012 leading sections with piezocomposite trailing control surfaces are performed at a fixed Reynolds number of 1000.
The effects of length and camber of the trailing control surface on the laminar aerodynamic characteristics are studied over a wide range of angle of attack.
It is shown that the flow behind the airfoil exhibits different features, including steady flow, periodic vortex shedding, and quasi-periodic vortex shedding for different configurations.
The transition between these wake states occurs at slightly smaller angles of attack compared to the nominal NACA 0012 airfoil.
While the drag coefficient remains close to each other at a fixed angle of attack, the lift coefficient of the hybrid airfoil is positively affected by the length and camber of the trailing control surface.
The mechanism of lift generation is examined by surface pressure distributions and a force element analysis.
It is revealed that with increased camber of the trailing control surface, the flow on both the suction and pressure sides of the airfoil are modified in a beneficial way to enhance lift. 
Increasing the length ratio only significantly modifies the flow near the aft section on the pressure side.
The results herein provide a laminar aerodynamic characterization of hybrid airfoils with trailing control surfaces, and could potentially aid the design of control techniques for next-generation small unmanned aerial vehicles.
\end{abstract}

\section*{Nomenclature}

{\renewcommand\arraystretch{1.0}
	\noindent\begin{longtable*}{@{}l @{\quad=\quad} l@{}}
		$\alpha_i$  & incidence of the leading airfoil \\
		$\alpha$ &    angle of attack of the airfoil\\
		$C_p$& pressure coefficient \\
		$C_d, C_l$ & drag and lift coefficients \\
		$c_{LS}$   & chord of the leading rigid segment of the airfoil\\
		$c_{TS}$   & chord of the trailing morphing segment of the airfoil\\
		$\delta$ & camber of the trailing morphing segment of the airfoil \\
		$\boldsymbol{e_y}$ & unit vector pointing in the $y$ direction\\
		$F_d, F_l$ & drag and lift forces \\
		$\nu$ & kinematic viscosity\\
		$L_e$ & lift element\\
		$l_{TS}$ & length of the trailing control surface\\
		$\boldsymbol{n}$ & wall normal unit vector\\
		$\omega_z$ & spanwise vorticity\\
		$p$ & pressure\\
		$\phi_L$ & auxiliary potential function for lift element analysis \\
		$r_C$ & camber ratio\\
		$Re$ & Reynolds number\\
		$r_L$ & length ratio\\
		$\rho$ & density\\
		$\boldsymbol{u}$ & velocity vector\\
		$U_{\infty}$ & freestream velocity\\
		$u_x, u_y$ & streamwise and crossflow velocity components\\
		$x, y$ & streamwise and crossflow coordinates\\
\end{longtable*}}

\section{Introduction}
\label{sec:intro}
In recent years, the use of unmanned aerial vehicles (UAVs) for remote observation and aerial photography is gaining popularity with the decreased size and weight of sensor, video, and communication devices.
To design these applications, fundamental understanding of flight at Reynolds number of $10^{2}-10^{5}$ becomes important.
Various research on this topic have been carried out to examine the Reynolds number effects \citep{dickinson1993unsteady,mueller2001fixed,winslow2018basic,rossi2018multiple}, wing shape effects \citep{pelletier2000low,sunada2002comparison,winslow2018basic}, end boundary effects \citep{torres2004low,kunihiko2009three,zhang2020formation,zhang2020laminar}, and so on.
At low Reynolds numbers, the flow over lifting bodies suffer from enhanced viscous effects, which make the flow more susceptible to separation, and limit the aerodynamic performance of the aircraft \citep{lissaman1983low,mueller2003aerodynamics,eldredge2019leading}.
Developing effective control techniques for low-Reynolds-number flows to counter such unfavourable flow conditions could lead to the design of next-generation UAVs.

Morphing wing technology is a method to improve aerodynamic efficiency and flight performance by changing the wing shape during flight \citep{sofla2010shape,cattafesta2011actuators,barbarino2011review,li2018review}. 
For large commercial aircraft and civil aviation helicopters, the modification of the wing configurations is usually achieved by movable flaps, slats, tabs. 
These devices require complex servomechanisms consisting of a large number of moving mechanical parts such as hinges, linkages, gears and bearings \citep{dimino2017morphing}.
However, such complex actuation systems could be very difficult to implement in small UAVs.

\begin{figure}[!htb]
\centering
\includegraphics[width=0.99\textwidth]{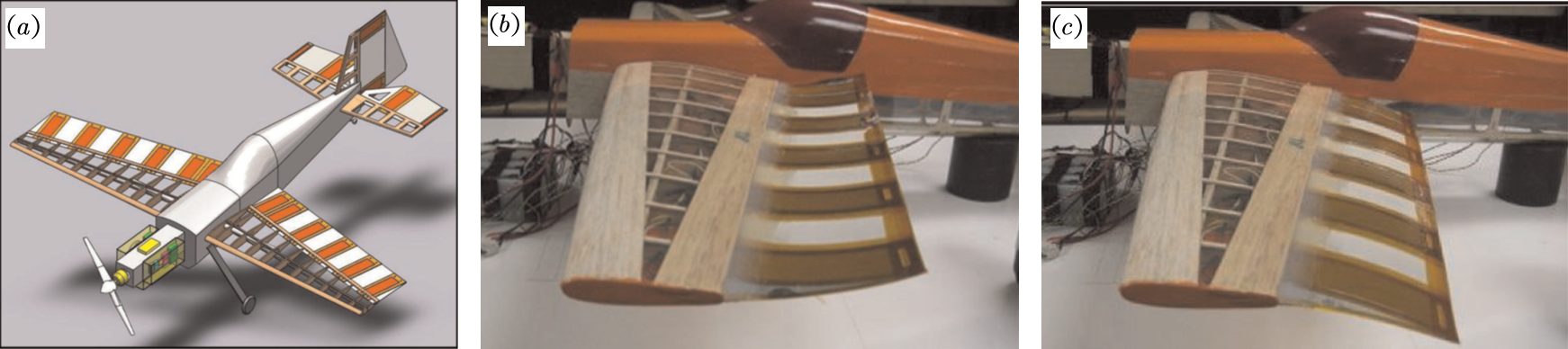}
\caption{(a) Model of the morphing aircraft with piezocomposite surfaces by \citet{bilgen2013novel}. (b) peak negative actuation and (c) peak positive actuation.}
\label{fig:prototype}
\end{figure}

Leveraging advances in material science, a so-called solid-state morphing control surface, which integrates the actuator with the structure and eliminates conventional mechanism, can substantially reduce  mechanical complexity, and is a promising solution to the problem outlined above \citep{giurgiutiu2000review,sun2016morphing,bilgen2009macro,bilgen2010macro,bilgen2010novel,bilgen2011theoretical,bilgen2013implementation,bilgen2014piezoceramic}.
A successful application of this technique, as shown in figure \ref{fig:prototype}, has been demonstrated by \citet{bilgen2013novel}.
In this work, the authors attached bimorph piezocomposite actuators to the trailing edge of a small model-scale aircraft.
This trailing control surface uses the Macro-Fiber Composite, which is a type of piezo-electric actuator that offers structural flexibility and high actuation authority.
In flight tests, fully morphing flights were achieved, demonstrating the potential of this concept.
Inspired by this research, similar trailing bimorph control configurations were developed by \citet{ohanian2011piezoelectric,probst2012smart}, and \citet{ohanian2012piezoelectric}.

With the morphing of trailing control surfaces, the wings can exhibit various kinds of shapes, providing a wide range of control. 
On the aerodynamics side, a preliminary study on the aerodynamic characteristics of this hybrid airfoil configuration has been conducted by \citet{bilgen2013novel}, where the authors analyzed the lift and drag coefficients using the panel method. 
While these theoretical analyses provided some insight into the aerodynamic performance of the morphing wing, the accuracy of these aerodynamic data are limited to relatively high speed flows, as the potential flow model can not account for the flow separation effects that are prevalent at low Reynolds numbers.
In addition, the mechanism of lift generation for these hybrid airfoils were not discussed in previous studies.

As a preliminary step to the design and control of a UAV utilizing piezocomposite trailing surfaces, a thorough understanding of the low-Reynolds-number aerodynamics of the hybrid airfoil configuration as shown in figure \ref{fig:prototype} is indispensable.
The aim of this study is to provide new insights into the wake dynamics and aerodynamic performance of hybrid airfoils comprised of a rigid leading section with a piezocomposite trailing control surface at low Reynolds number.
To this end, the two-dimensional laminar flow over the hybrid airfoil with different length and camber of the trailing control surface is studied by direct numerical simulations over a wide range of angle of attack at a fixed Reynolds number of 1000.
In what follows, the concept of the hybrid solid-state airfoil is illustrated in \S \ref{sec:concept}. 
The computational model and its validation are presented in \S \ref{sec:computation}.
In \S \ref{sec:results}, the results from the parametric analysis are discussed from the perspectives of wake features and aerodynamic forces. 
In addition, the mechanism of lift enhancement by the trailing surface morphing is explained.
Concluding remarks are offered in \S \ref{sec:conclusions}.
These results should provide new knowledge about the low-Reynolds-number aerodynamic behavior of this novel hybrid airfoil family, and could potentially aid the development of next-generation small unmanned aerial vehicles.

\section{The solid-state airfoil concept}
\label{sec:concept}
\begin{figure}[!htb]
\centering
\includegraphics[scale=0.16]{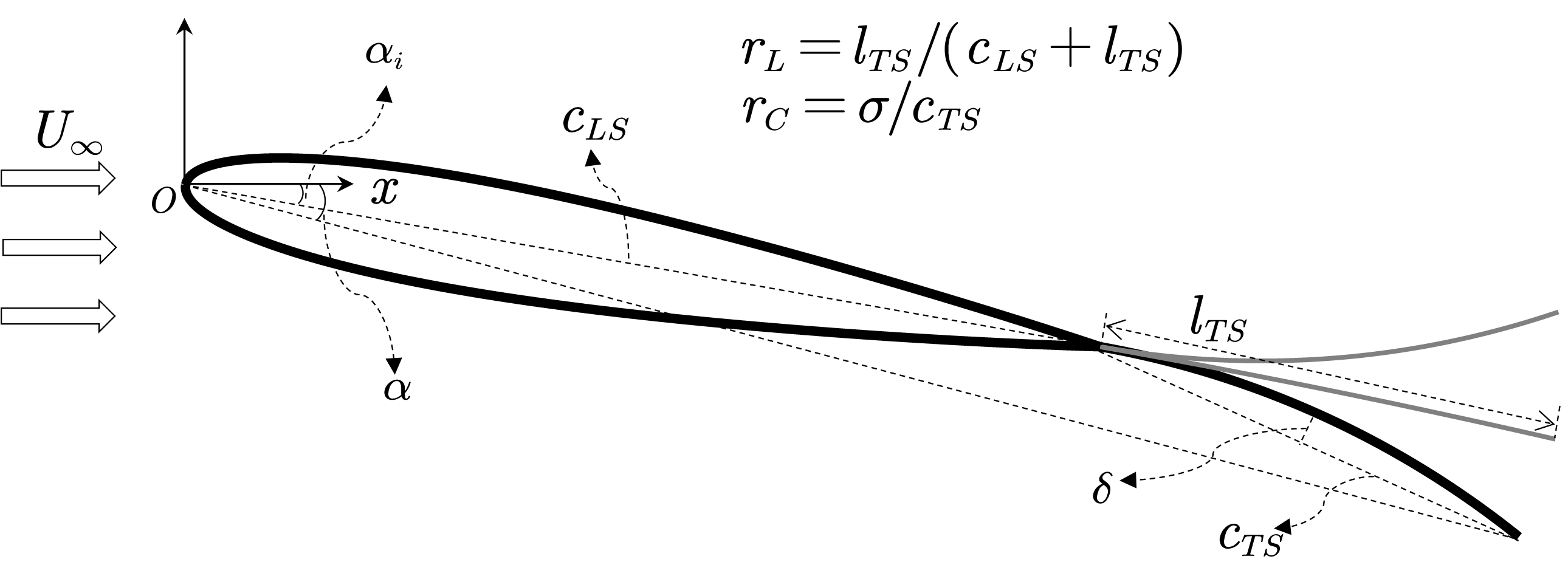}
\caption{Design of the solid-state airfoil with piezocomposite trailing section.}
\label{fig:description}
\end{figure}

The design of the solid-state hybrid airfoil is illustrated in figure \ref{fig:description}.
The leading section is assumed to be a rigid NACA 0012 airfoil which has a chord length of $c_{LS}$, and angled to the freestream $U_{\infty}$ at an incidence $\alpha_i$.
The trailing morphing segment is attached to the end of the leading section, and has a length of $l_{TS}$, resulting in a trailing length ratio of $r_L=l_{TS}/(l_{TS}+c_{LS})$.
The thickness of the trailing section is idealized to be zero in the analysis.
Without actuation, the trailing control surface remains straight, and extends along the chord line of the leading segment.
When actuated, the trailing section deflects to an arc with a camber ratio $r_C=\delta/c_{TS}$, where $\delta$ and $c_{TS}$ are the camber and chord of the deflected control surface. 
The trailing section is assumed to be a piezocomposite bimorph, and can deflect downward ($\delta>0$) with positive control input, and bend upward ($\delta <0$) when the control input is negative.
Here, the Reynolds number, which is fixed at 1000, is defined as $Re=U_{\infty}(c_{LS}+l_{TS})/\nu$, where $U_{\infty}$ is the freestream velocity, and $\nu$ is the kinematic viscosity of the fluid.
The angle of attack of the hybrid airfoil $\alpha$ is determined by the leading incidence angle $\alpha_i$, length ratio $r_L$ and camber ratio $r_C$.
In what follows, the length quantities are reported in their dimensionless forms by normalizing with $l_{TS}+c_{LS}$, velocity with $U_{\infty}$, and time with $(l_{TS}+c_{LS})/U_{\infty}$.

\section{Computational fluid dynamics model}
\label{sec:computation}

\begin{figure}[!htb]
\centering
\includegraphics[width=0.8\textwidth]{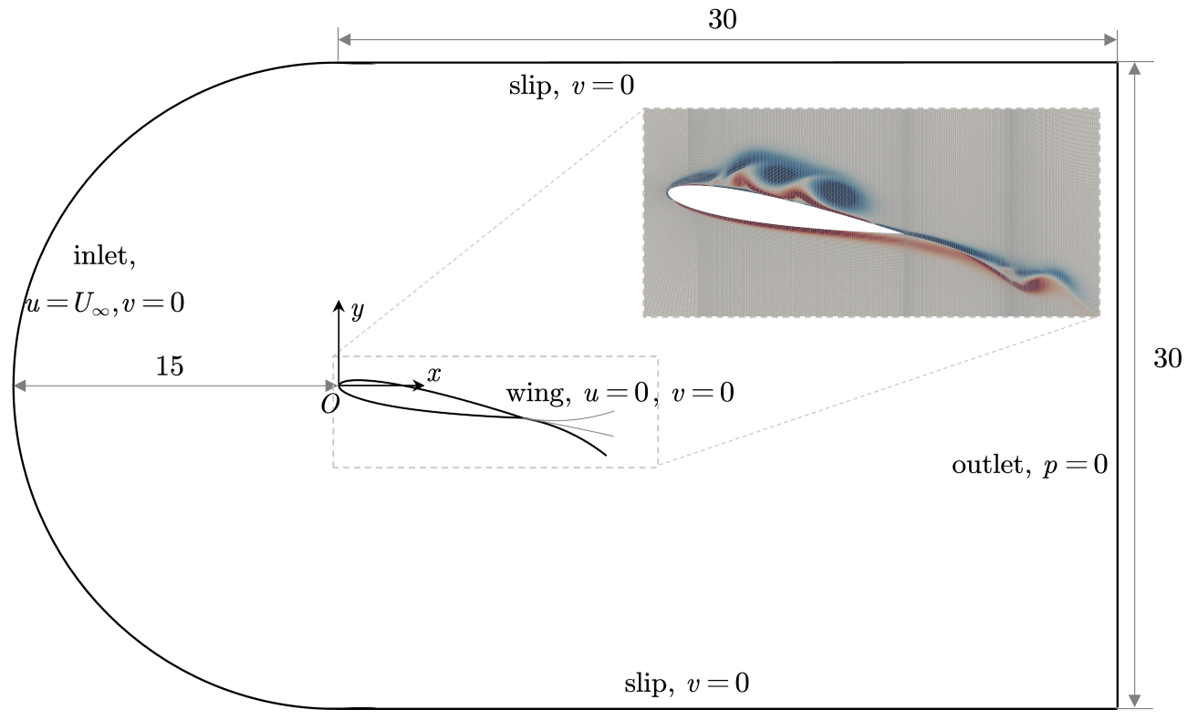}
\caption{Computational domain and boundary conditions. The inset shows grid resolution in the vicinity of the airfoil overlaid on the vorticity field ($\omega_z=\partial u_x/\partial y- \partial u_y/\partial x$) at $Re=1000$. Shown is $\omega_z\in\left[-10,10\right]$, with blue color represents negative value and red positive.}
\label{fig:mesh}
\end{figure}

The open-source CFD toolbox OpenFOAM-v2.4.x \citep{weller1998tensorial}, which employs the finite-volume method, is used for simulating the two-dimensional flow over airfoil designs. 
The unsteady flow solver \textit{pimpleFoam} is used with second-order-accurate spatial and temporal schemes to solve the incompressible Navier-Stokes equations.
The computational domain and boundary conditions are presented in figure \ref{fig:mesh}.
With the leading-edge of the airfoil placed at $(x,y)=(0,0)$, the computational domain covers $(x,y)\in[-15,30]\times[-15,15]$, where $x$ and $y$ are the streamwise and crossflow coordinates. 
This results in a maximum blockage ratio of 1\% for the case $(r_L,r_C,\alpha)=(0.5,0.1,15^{\circ})$. 
The inlet is prescribed with freestream velocity $\boldsymbol{u}=(U_{\infty},0)$. 
The slip boundary condition is used for the side boundaries.
The airfoil surface is treated with a no-slip boundary condition. The zero gradient condition is applied to the outlet, where a reference pressure $p=0$ is specified. 
The computational domain is discretized with a C-type grid with the mesh refined in the vicinity of the airfoil, as shown in the inset in figure \ref{fig:mesh}. 
The first layer of the mesh from the airfoil surface is $\Delta y_0=0.004$.
The time-step is set to be $\Delta t=0.001$, which results in a Courant–Friedrichs–Lewy (CFL) number of less than one.
The simulations are started with uniform flow.
After the initial transients are flushed out of the computational domain, the flow statistics are collected with over 100 convective time units to ensure convergence.

\begin{figure}[!htb]
\centering
\includegraphics[width=0.5\textwidth]{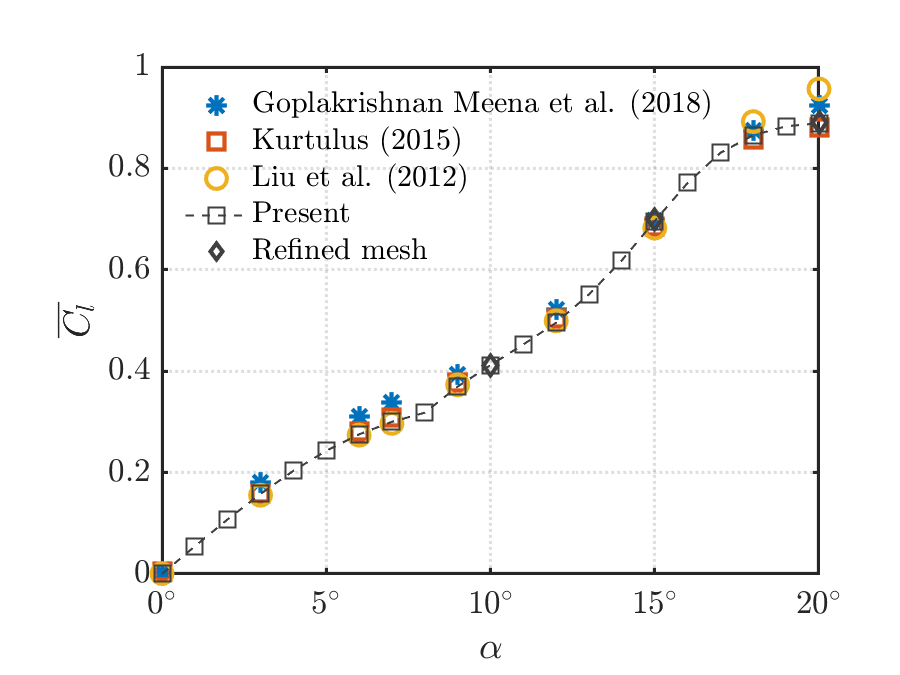}
\caption{Comparison of lift coefficient for a NACA 0012 airfoil at $Re=1000$.}
\label{fig:validation}
\end{figure}

To validate the accuracy of the computational setup, figure \ref{fig:validation} compares current prediction of lift coefficient for a NACA 0012 airfoil at $Re=1000$ with past studies \citep{liu2012numerical,kurtulus2015unsteady,gopalakrishnan2017airfoil}. The same plot also presents results obtained from a refined mesh and time-step ($\Delta y_0=0.002$ and $\Delta t=0.0005$).
The lift and drag coefficients are defined as 
\begin{equation}
C_l = \frac{F_l}{\frac{1}{2}\rho U_{\infty}^2 (c_{LS}+l_{TS})} \quad \textrm{and} \quad C_d = \frac{F_d}{\frac{1}{2}\rho U_{\infty}^2 (c_{LS}+l_{TS})},
\end{equation}
where $F_l$ and $F_d$ are the lift and drag forces on the two-dimensional airfoil.
The lift coefficient obtained from the current simulations is in good agreement with the data from literature. 
It is also seen that the refined mesh does not significantly affect the results.
These observations demonstrate the accuracy of the computational method.

\section{Parametric aerodynamic analyses}
\label{sec:results}
In this section, a parametric study covering length ratio $r_L \in [0, 0.5]$, camber ratio $r_C\in [-0.1,0.1]$ and leading segment incidence angle $\alpha_i\in[0,15]$, at a fixed Reynolds number $Re=1000$ is performed.
The flow features are discussed first, followed by a classification of wake regimes in the $r_L$-$r_C$-$\alpha_i$ space. 
Next, the aerodynamic forces are analyzed in detail.
Finally, the mechanism for lift generation is explored from two perspectives: surface pressure distribution and force element analysis.

\subsection{Wake dynamics}
\label{sec:wake}
\begin{figure}[htb!]
\centering
\includegraphics[width=0.99\textwidth]{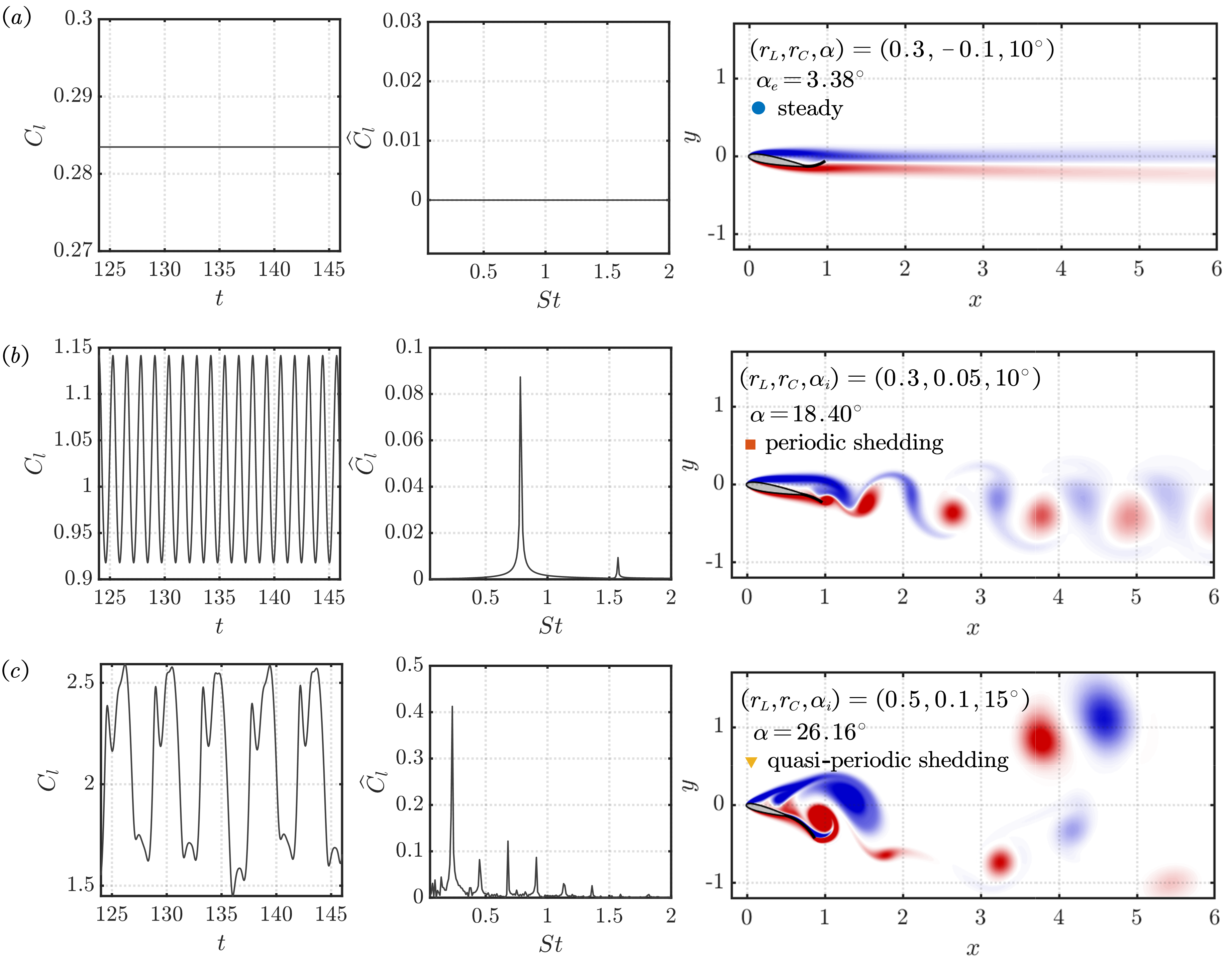}
\caption{Representative wake types visualized by (from left to right) lift coefficients, lift amplitude spectrum and vorticity fields ($\omega_z\in[-10,10]$, with blue and red representing negative and positive vorticity values). ($a$) Steady flow $(r_L,r_C,\alpha_i)=(0.3,-0.1,10^{\circ})$; ($b$) periodic shedding $(r_L,r_C,\alpha_i)=(0.3,0.05,10^{\circ})$; ($c$) quasi-periodic shedding $(r_L,r_C,\alpha_i)=(0.5,0.1,15^{\circ})$.}
\label{fig:liftAndWake}
\end{figure}

The wake of the airfoil exhibits different dynamical features depending on the length ratio $r_L$, camber ratio $r_C$ of the trailing control surface, and the incidence angle of the leading segment $\alpha_i$.
The representative wake types are presented in figure \ref{fig:liftAndWake}, which shows the lift coefficient, lift spectrum, and spanwise vorticity field.
For the case $(r_L,r_C,\alpha_i)=(0.3,-0.1,10^{\circ})$, which is associated with a low angle of attack $\alpha=3.38^{\circ}$, the flow around the airfoil is characterized by steady continuous vortex sheets with constant lift.
At $(r_L,r_C,\alpha_i)=(0.3, 0.5, 10^{\circ})$, the flow is characterized by shedding of vortex pairs in the wake, as shown in figure \ref{fig:liftAndWake}$(b)$.
The lift coefficient is periodic with respect to time, with a single dominant frequency and its superharmonic in its spectrum.
This type of periodic shedding is typical for airfoils with moderate angle of attack ($\alpha=18.4^{\circ}$ in this case).
When the effective angle of attack reaches a high value, such as figure \ref{fig:liftAndWake}($c$), two distinct pairs of counter-rotating vortices are formed in the wake. The weaker pair moves along the freestream direction, and the stronger pair possesses an upward motion as it convects towards the wake.
The lift spectrum for this case features several peaks.
For the case shown here, the quasi-periodic lift history is also interspersed with stochastic variations, suggesting the onset of irregular vortex shedding \citep{rossi2018multiple}.
In the case where $\alpha_i$ is small and the camber ratio is large, the airfoil can arrive at negative angle of attack, resulting in negative lift. 
These flow features have also been reported in past studies for lifting surfaces at similar Reynolds numbers. \citep{kurtulus2015unsteady,gopalakrishnan2017airfoil,rossi2018multiple,menon2019aerodynamic}.

\begin{figure}[htb!]
\centering
\includegraphics[width=0.99\textwidth]{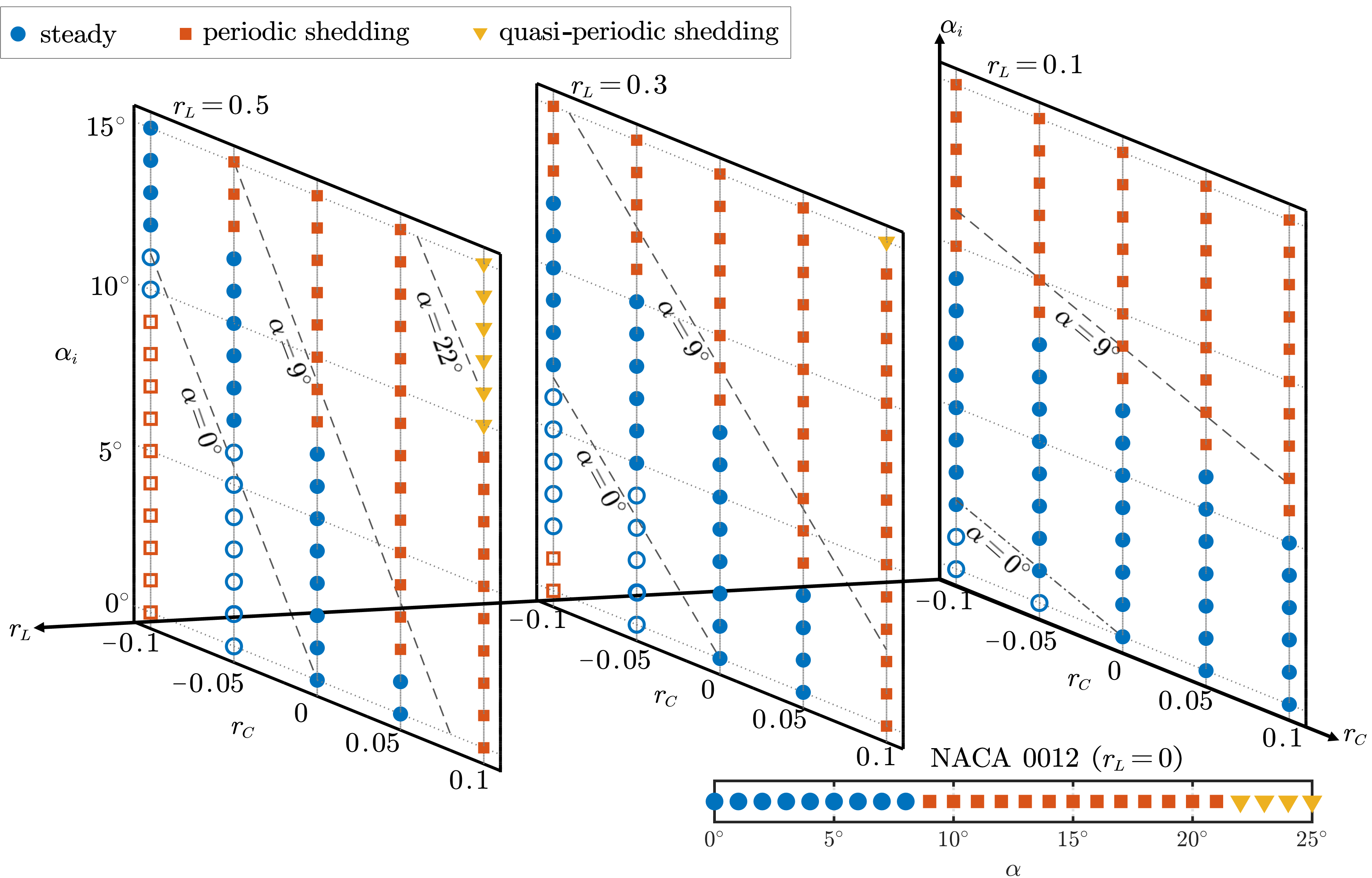}
\caption{Classification of flow features in the $r_L$-$r_C$-$\alpha_i$ space. Blue cirle, orange square, and yellow triangle represent steady flow, periodic shedding, and quasi-periodic shedding. Hollow marks represent cases with negative lift force. The flow regimes for a nominal NACA 0012 airfoil (i.e., $r_L=0$) at $Re=1000$ is shown in the bottom right. Contour lines of angles of attack $\alpha=0^{\circ}, 9^{\circ}$ and $22^{\circ}$ are shown in dashed lines.}
\label{fig:regime}
\end{figure}

A classification of the flows in the $r_L$-$r_C$-$\alpha_i$ space is presented in figure \ref{fig:regime}. 
For flow over a nominal NACA 0012 airfoil at $Re=1000$, the wake transitions from steady flow to periodic shedding at $\alpha_i=9^{\circ}$, and then transitions to quasi-periodic shedding at $\alpha_i=22^{\circ}$. 
The contour lines of $\alpha$ at these two transitional angles of attack, plus $\alpha=0^{\circ}$ (boundary for positive and negative lift) are also depicted in the $r_C$-$\alpha_i$ planes for each $r_L$ as references.
At $r_L=0.1$, the wake exhibits steady flow for cases with low $\alpha_i$, and periodic shedding for high $\alpha_i$. The transitional angle of attack between these two flow states decreases with increasing camber ratio $r_C$ in an almost linear manner. 
This boundary is slightly lower than the $\alpha=9^{\circ}$ contour line, indicating that the use of a thin trailing surface promotes the unsteady vortex shedding compared to a nominal NACA 0012 airfoil.
Negative lift (denoted by hollow marks) with steady flow is shown in the bottom left corner of the plot, and is well predicted by the $\alpha=0^{\circ}$ contour line.

At $r_L=0.3$ and $r_L=0.5$, negative lift associated with both steady flow and periodic shedding is achieved at wider $\alpha_i$ ranges when the trailing control surface is actuated to bend upward (negative $r_C$). 
As with the case for $r_L=0.1$, the contour line of $\alpha=0^{\circ}$ fits well with the negative-positive lift boundaries. 
The transition $\alpha_i$ between the steady flow and periodic shedding are still below the $\alpha=9^{\circ}$ contour line, although they are no longer linear in the $r_C$-$\alpha_i$ space.
With high angles of attack and high camber ratio (top right corner of the $\alpha_i$-$r_C$ planes), flow with quasi-periodic shedding can be observed. 
Again, the fact that these quasi-periodic flows occur before the angle of attack reaches $\alpha=22^{\circ}$ suggests the importance of airfoil shape on the wake dynamics of low-$Re$ airfoils.

\subsection{Aerodynamic forces}

\subsubsection{Force coefficients as a function of leading section incidence angle}
\begin{figure}[htb!]
\centering
\includegraphics[width=0.99\textwidth]{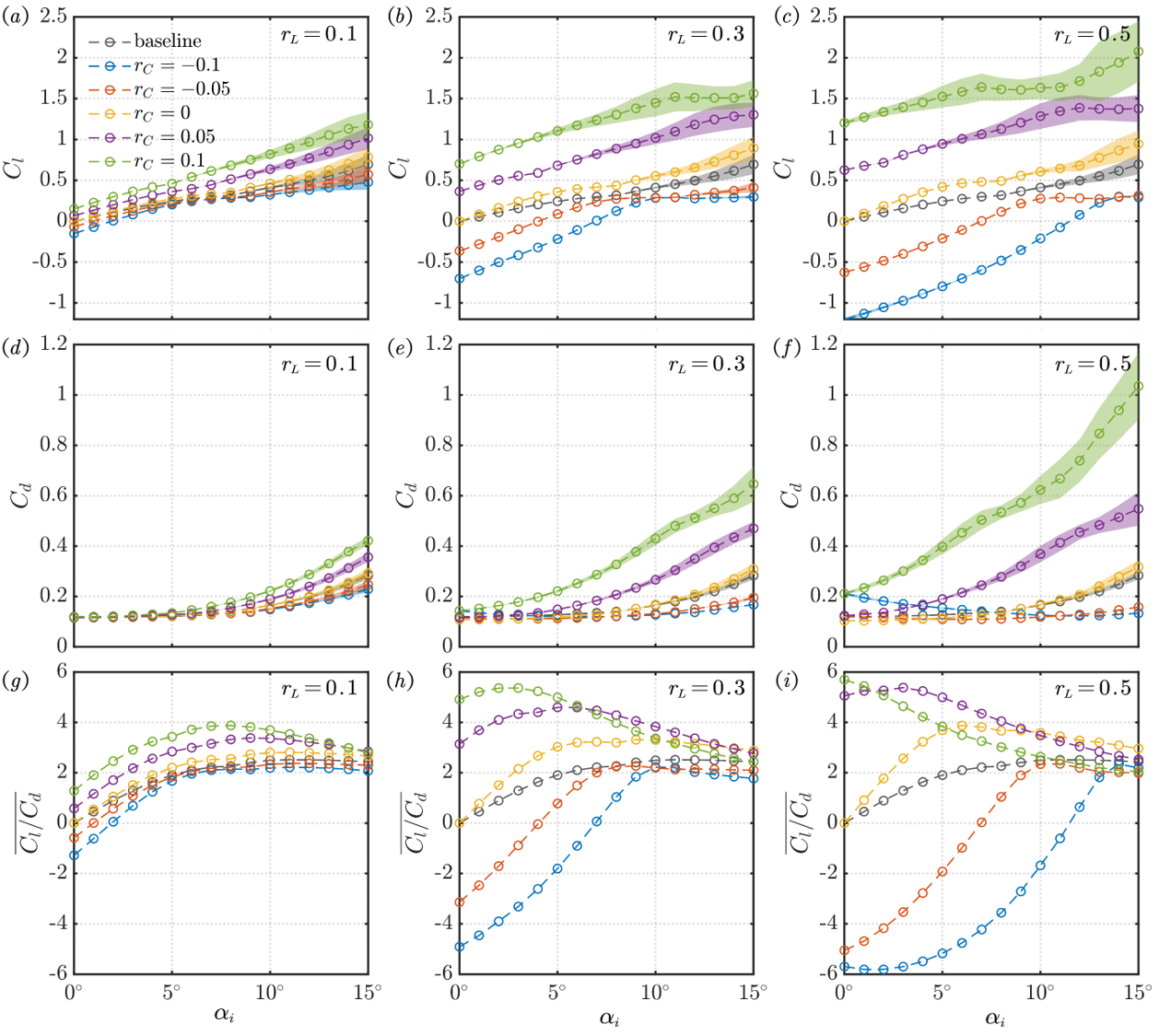}
\caption{Lift coefficient (first row), drag coefficient (second row) and time-averaged lift-to-drag ratio (last row) for cases with $r_L=0.1$ (left column), $r_L=0.3$ (middle column) and $r_L=0.5$ (right column). In the lift and drag coefficients plots, the dash lines represent the time-average values, and the shaded regions indicate the rms values of the coefficients.}
\label{fig:forceCoeffs}
\end{figure}

The lift, drag coefficients, and the lift-to-drag ratio as a function of the leading section incidence $\alpha_i$ are compiled in figure \ref{fig:forceCoeffs} for cases with different length and camber ratios.
As is typical for low-Reynolds-number flow \cite{gopalakrishnan2017airfoil,menon2019aerodynamic,zhang2020formation}, the lift coefficient of the hybrid airfoil generally increases with incidence $\alpha_i$, except for a few plateaus in the $C_l$-$\alpha_i$ curves observed in cases with large length ratio ($r_L=0.3$ and 0.5).
With increasing length ratio, the airfoil experiences extended variation of lift by varying the camber, suggesting higher control efficacy.
However, with high $r_L$, the wake is more prone to transitioning to unsteady periodic shedding or quasi-periodic shedding, which invites larger fluctuating lift. 

The drag coefficient shown in figure \ref{fig:forceCoeffs}$(d-f)$ generally exhibits a quadratic-like growth with $\alpha_i$.
Exceptions are observed for hybrid airfoils with negative angle of attack (large length ratio and negative camber), where the drag coefficient decreases with the incidence angle.
With increasing length of the trailing control surface, the hybrid airfoil suffers from significantly increased drag as the camber turns from negative to positive.
The lift-to-drag ratio, as an indicator of aerodynamic efficiency generally increases with the camber ratio.
At $r_L=0.1$, the $\overline{C_l/C_d}$-$\alpha_i$ curves pose a similar trend with the baseline NACA 0012 airfoil. 
With larger length ratios, the lift-to-drag ratio of the hybrid airfoil varies significantly among cases with different camber ratio at low incidence, and converge to $\overline{C_l/C_d}\approx 2$ at high incidence.

\subsubsection{Force coefficients as a function of angle of attack}
\begin{figure}[htb!]
\centering
\includegraphics[width=0.99\textwidth]{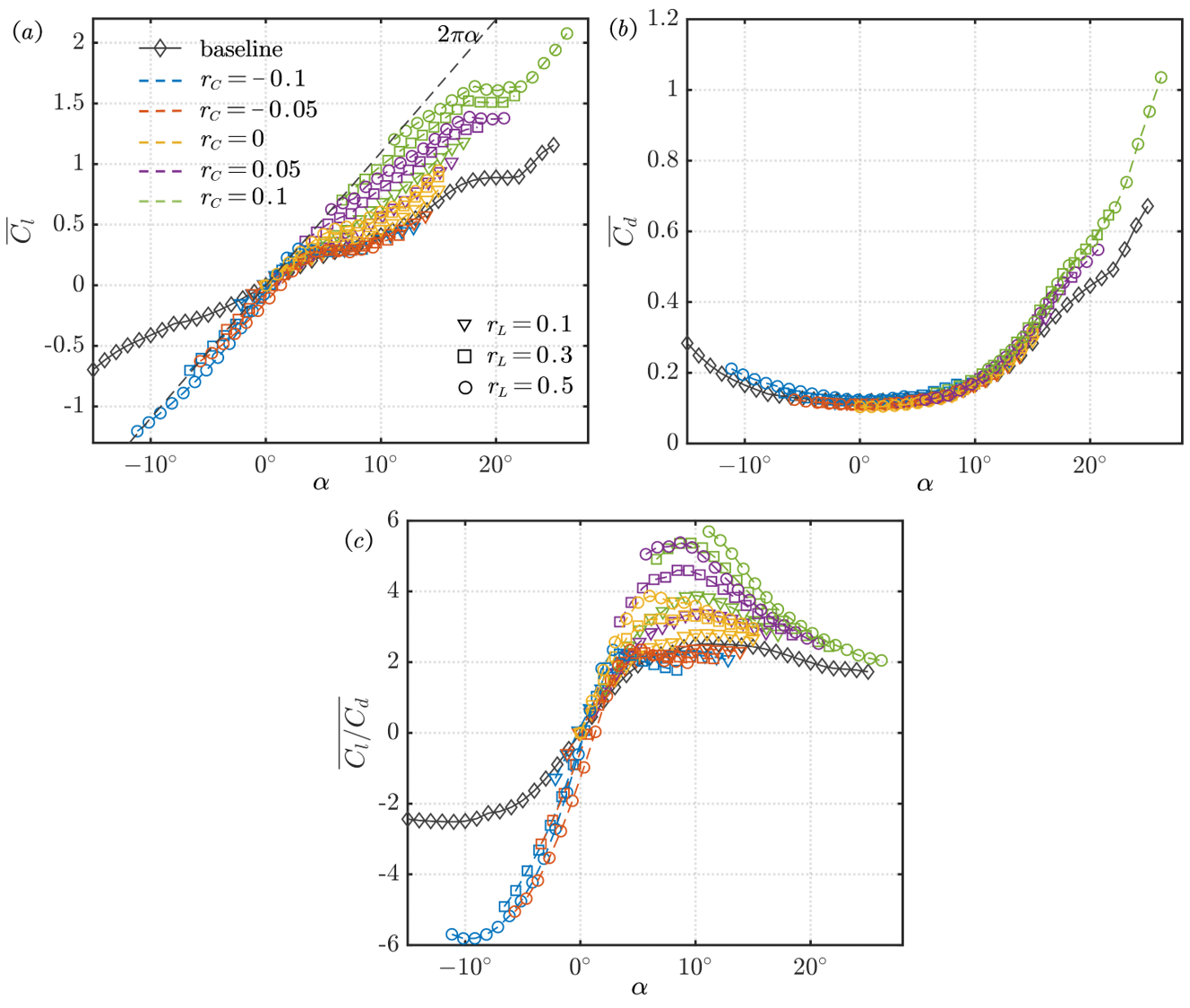}
\caption{($a$) Lift coefficient, ($b$) drag coefficient, and ($c$) lift-to-drag ratio as a function of angle of attack $\alpha$. In the insets, the force coefficients at a constant angle of attack $\alpha=10^{\circ}$ are plotted for varying $r_C$ at a fixed $r_L=0.3$ (red) and varying $r_L$ at a fixed $r_C=0.05$ (blue). The shaded region indicates the root-mean-squared values.}
\label{fig:forceCoeffEff}
\end{figure}

\begin{figure}[htb!]
\centering
\includegraphics[width=0.99\textwidth]{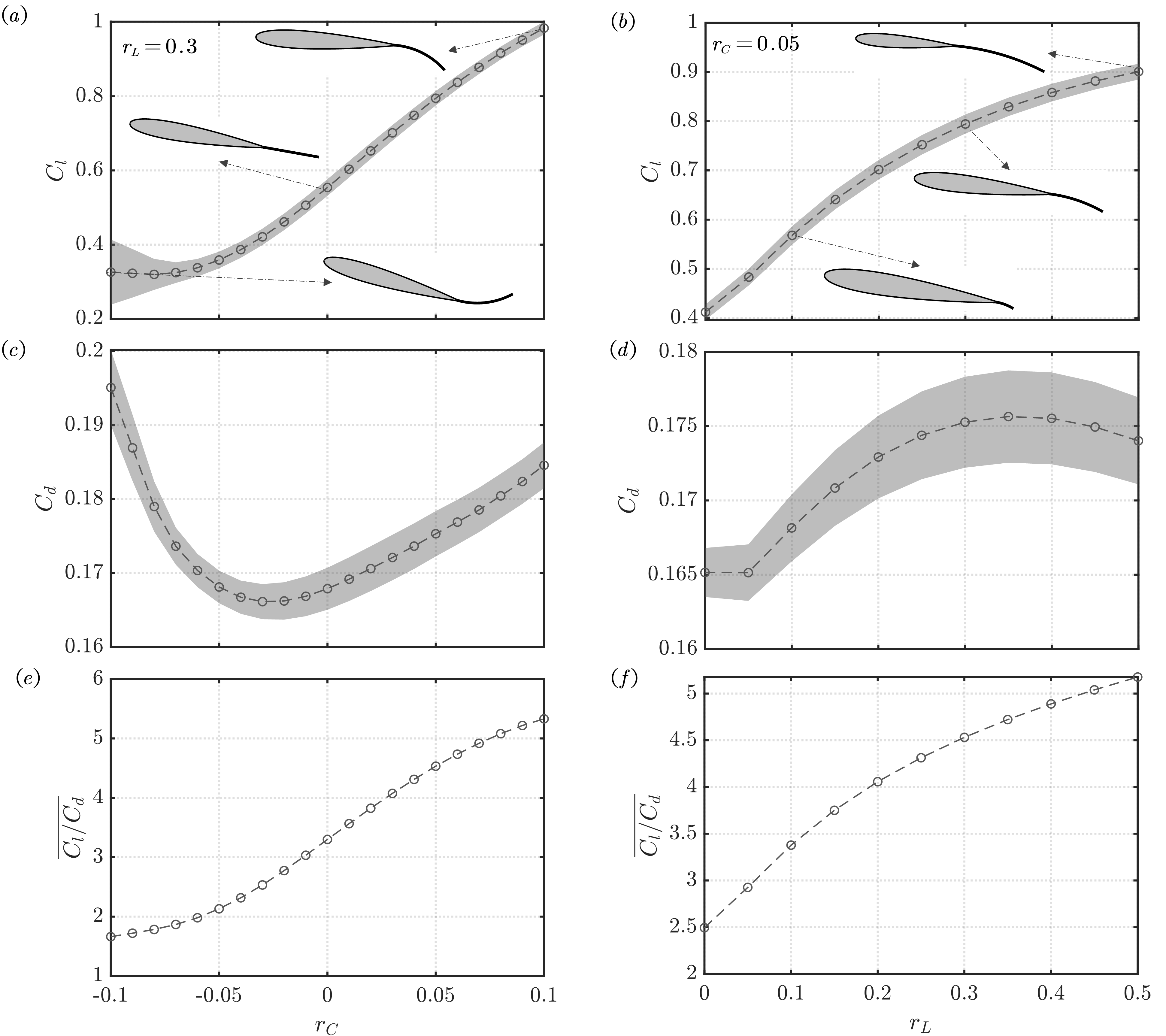}
\caption{Lift coefficient (top), drag coefficient (middle), and lift-to-drag ratio (bottom) for a fixed angle of attack $\alpha=10^{\circ}$. ($a,c,e$) fixed $r_L=0.3$ and varying $r_C$ from -0.1 to 0.1; ($b,d,f$) fixed $r_C=0.05$ and varying $r_L$ from 0 to 0.5. Mean values are shown in dashed lines, and the shaded regions indicate the rms values. The configurations of the airfoils are plotted as insets in the top figures for selected cases.}
\label{fig:AoAe10}
\end{figure}

The aerodynamic force coefficients as a function of the angle of attack $\alpha$ are compiled in figure \ref{fig:forceCoeffEff} to better understand the data presented in figure \ref{fig:forceCoeffs}.
At the same angle of attack, the lift of hybrid airfoils of different configurations can vary significantly.
In general, the lift is positively related with both the length and camber of the trailing control surface.
This is more clearly shown in figure \ref{fig:AoAe10}$(a)$ and $(b)$, in which the lift increases monotonically with $r_C$ and $r_L$ at a fixed angle of attack $\alpha=10^{\circ}$.
The mechanism of lift generation will be further discussed in \S \ref{sec:mechanism}.

The drag coefficients of all cases collapse well onto each other for $-10^{\circ}\lesssim\alpha\lesssim 15^{\circ}$, as observed from figure \ref{fig:forceCoeffEff}($b$).
A closer examination of drag coefficient from figure \ref{fig:AoAe10}($c$) and $(d)$ reveals that for a fixed angle of attack $\alpha=10^{\circ}$, $C_d$ exhibits non-monotonic relationship with $r_C$ and $r_L$.
Nevertheless, the variation in drag with varying $r_C$ or $r_L$ is much smaller than the lift.
With increasing camber ratio, the angle of attack at which maximum lift-to-drag ratio is achieved shifts from $\alpha\approx 5^{\circ}$ for negative $r_C$ to $\alpha\approx 10^{\circ}$ for positive $r_C$.
The maximum $\overline{C_L/C_D}$ itself is positively related to both $r_C$ and $r_L$, as observed in the inset of figure \ref{fig:AoAe10}($e$) and ($f$).

\subsubsection{Lift derivative with respect to camber ratio}
The lift derivative with respect to the camber ratio at $r_C=0$ is presented in table \ref{tab:Clslope} for different sets of $(r_L, \alpha)$. 
This quantity serves as an indicator of the sensitivity of lift to the piezo-electric control at a particular operating condition of the hybrid airfoil, and is calculated as
\begin{equation}
	\displaystyle{\frac{\partial C_{l}}{\partial r_C}\Bigr|_{r_C=0} \approx \frac{C_{l}^{+}-C_{l}^{-}}{\Delta r_C}},
\end{equation}
in which $C_{l}^{\pm}$ is the lift coefficient computed at $r_C=\pm 0.01$ for fixed $\alpha_i$ and $r_L$. 
In the case of unsteady flow, the time-averaged force coefficient is used for calculating the slopes.
As observed in table \ref{tab:Clslope}, for airfoil with short trailing control surface ($r_L=0.1$), the lift derivative exhibit a monotonic increase with $\alpha_i$.
However, for airfoil with $r_L=0.3$ and 0.5, $\partial C_l/\partial r_C$ is also high $\alpha_i=0$. 
Nevertheless, the lift derivative increases with the length ratio $r_L$ across the studied $\alpha_i$, suggesting a higher sensitivity of lift for a longer trailing control surface.

\begin{table}[]
	\caption{\label{tab:Clslope} Lift derivative with respect to $r_C$}
	\centering
	\begin{tabular}{p{2cm}p{2cm}p{2cm}p{2cm}p{2cm}}
		\hline
		& $\alpha_i=0^{\circ}$ & $\alpha_i=5^{\circ}$ & $\alpha_i=10^{\circ}$ & $\alpha_i=15^{\circ}$ \\ \hline
		$r_L=0.1$ & 1.25 & 1.44 & 2.94 & 4.88 \\
		$r_L=0.3$ & 7.35 & 5.46 & 8.20 & 10.85 \\
		$r_L=0.5$ & 14.36 & 9.97 & 11.29 & 14.26 \\
		\hline
	\end{tabular}
\end{table}


\subsection{Mechanism for lift generation }
\label{sec:mechanism}

\begin{figure}[htb!]
\centering
\includegraphics[scale=0.45]{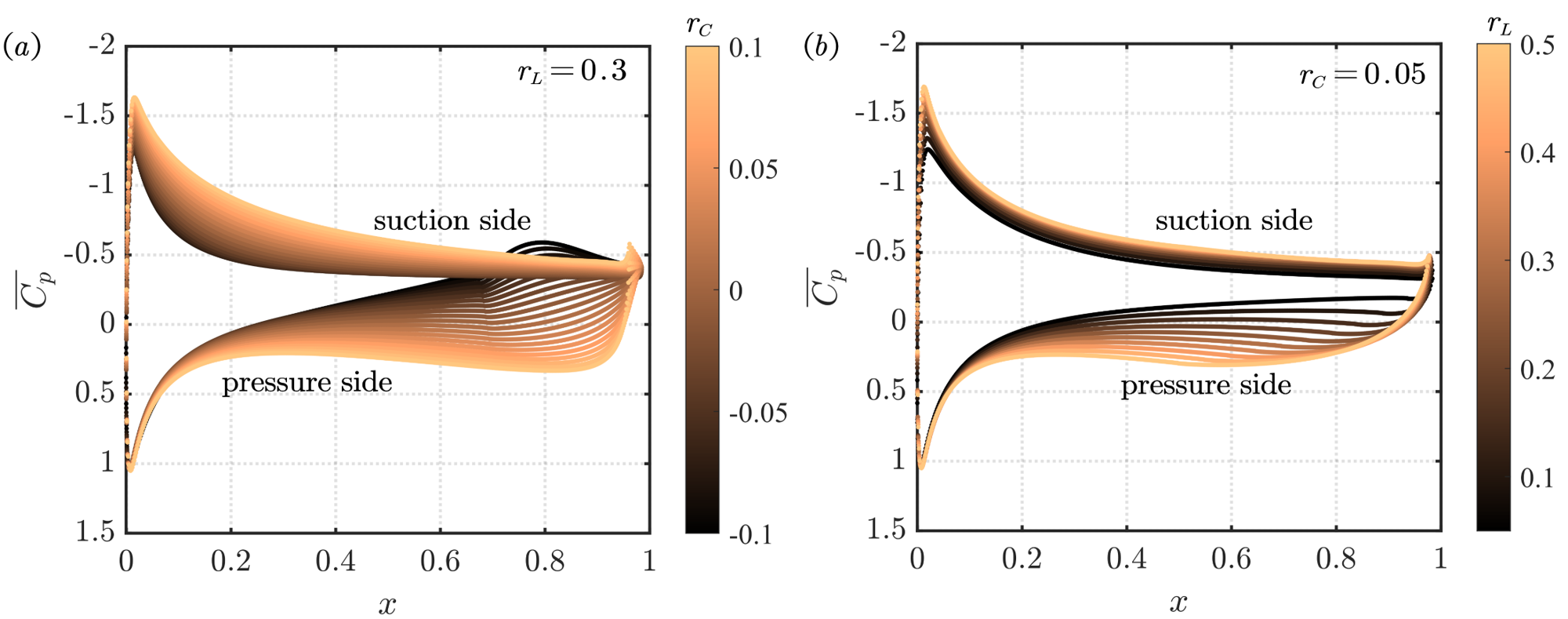}
\caption{Time-averaged pressure distributions on the airfoil surface at a fixed angle of attack of $10^{\circ}$. ($a$) Varying $r_C$ with fixed $r_L=0.3$ and ($b$) varying $r_L$ with fixed $r_C=0.05$.}
\label{fig:pressure}
\end{figure}

The mechanism of lift generation is explored for the airfoil at $\alpha=10^{\circ}$, considering varying $r_C$ at a fixed $r_L=0.3$, and varying $r_L$ at a fixed $r_C=0.05$.
The time-averaged pressure coefficient on the airfoil surface, which is defined as $C_p=(p-p_{\infty})/(\frac{1}{2}\rho U_{\infty}^2)$, is presented in figure \ref{fig:pressure}.
For airfoil with fixed $r_L=0.3$, increasing the trailing camber ratio $r_C$ affects the pressure distributions on both the suction and pressure sides in a favorable way that increases the lift.
On the suction side, the enhancement of the negative pressure is mostly observed near the fore part of the airfoil. 
On the pressure side, $\overline{C_p}$ is more strongly modified near the aft part. 
For airfoil with a fixed trailing camber ratio $r_C=0.05$, increasing $r_L$ significantly increases the pressure on the pressure side. 
In comparison, the negative pressure on the suction side is only slightly enhanced.

\begin{figure}[htb!]
\centering
\includegraphics[width=0.99\textwidth]{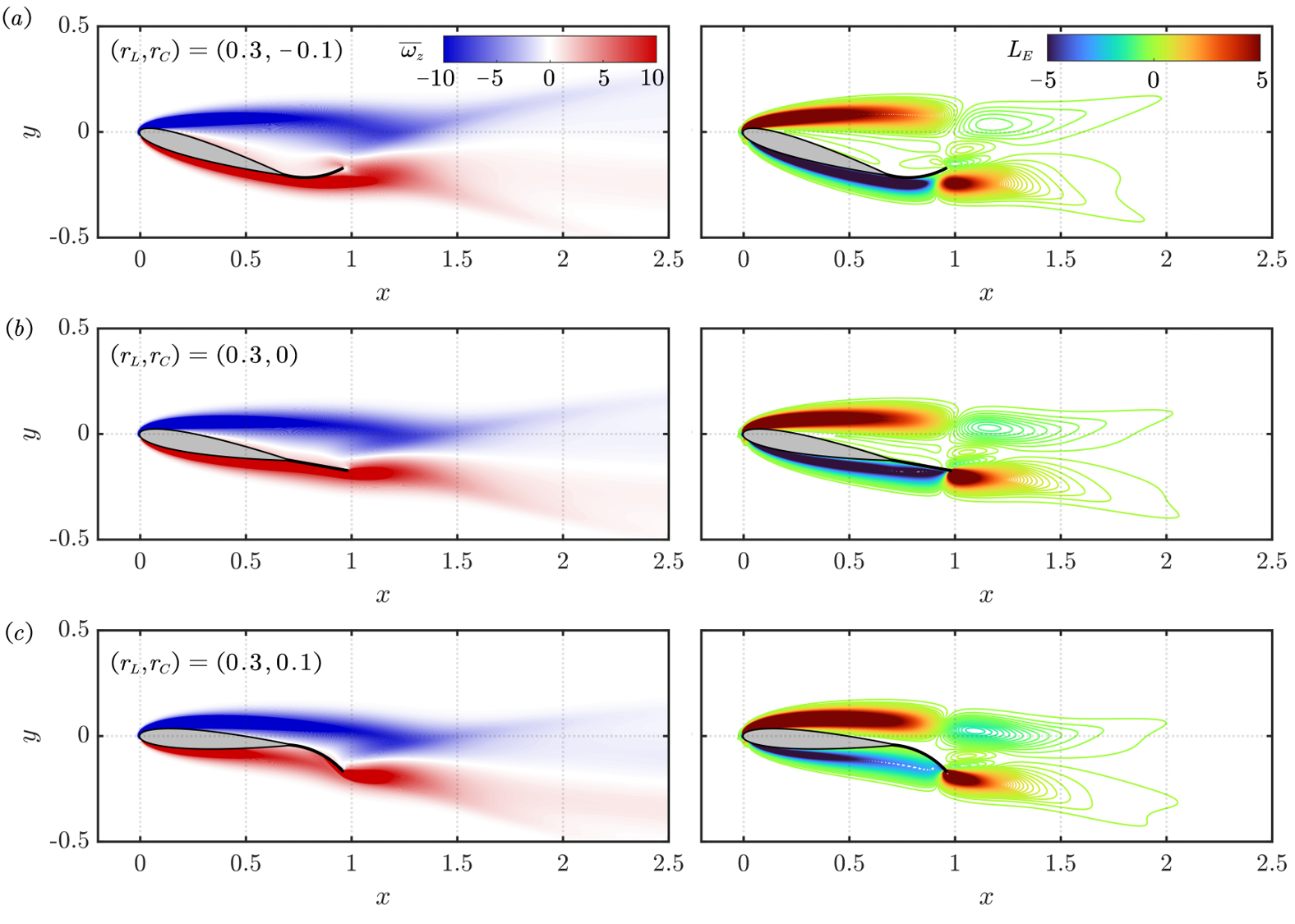}
\caption{Time-averaged vorticity fields (left) and lift element distributions (right) for cases with fixed $\alpha=10^{\circ}$ and $r_L=0.1$. $(a)$ $r_C=-0.1$; $(b)$ $r_C=0$ and $(c)$ $r_C=0.1$.}
\label{fig:rL03}
\end{figure}

\begin{figure}[htb!]
\centering
\includegraphics[width=0.99\textwidth]{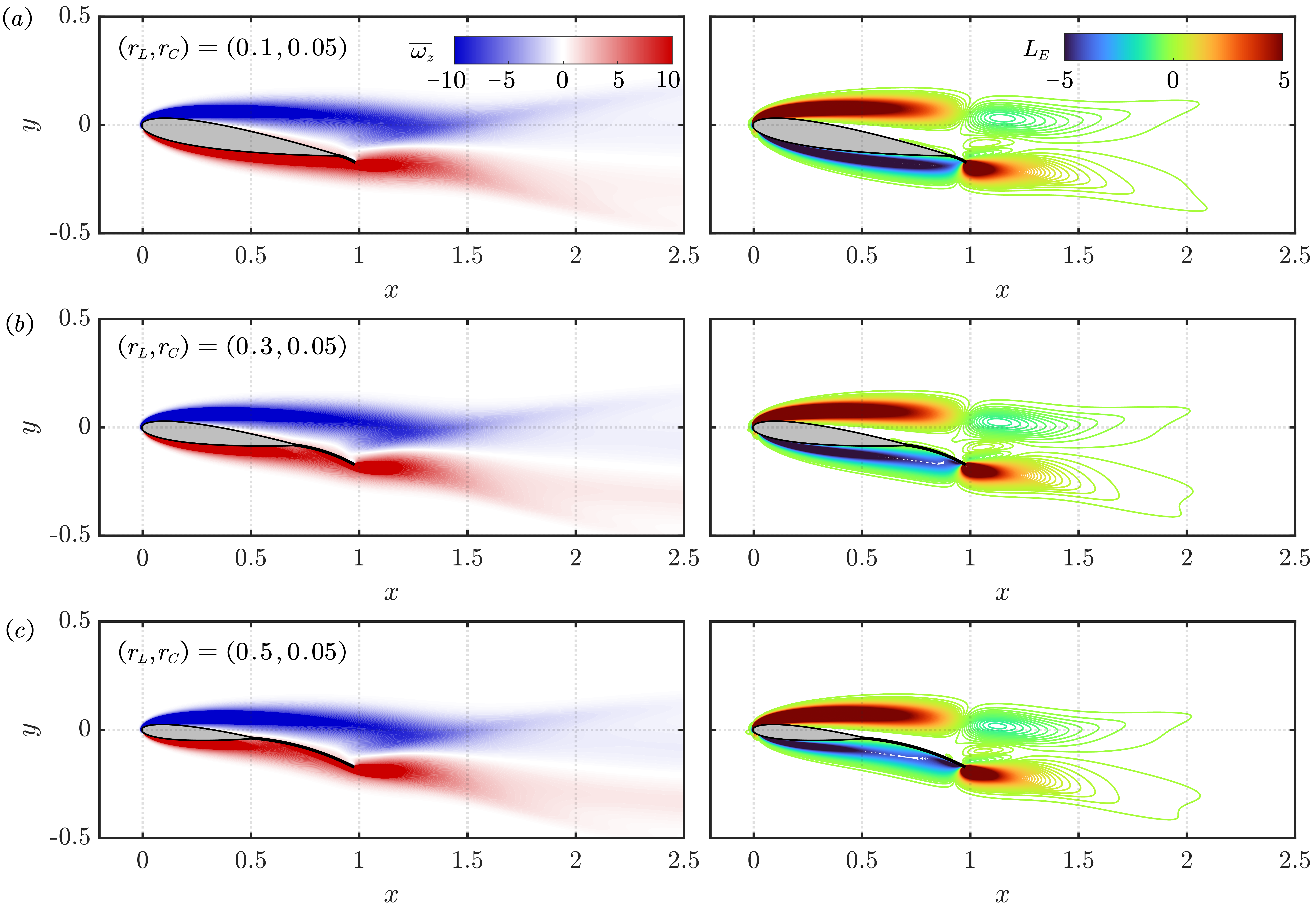}
\caption{Time-averaged vorticity fields (left) and lift element distributions (right) for cases with fixed $\alpha=10^{\circ}$ and $r_C=0.05$. $(a)$ $r_L=0.1$; $(b)$ $r_L=0.3$ and $(c)$ $r_L=0.5$.}
\label{fig:rC005}
\end{figure}

To gain further insight into the lift generation mechanism, the force element theory \citep{chang1992potential} is used to identify the flow structures that exert aerodynamic forces on the airfoil. 
For the lift force, we compute an auxiliary potential function $\phi_L$ (satisfying $\boldsymbol{\nabla}^2\phi_L=0$) with the boundary condition 
\begin{equation}
    -\boldsymbol{n}\cdot\boldsymbol{\nabla} \phi_L 
    = \boldsymbol{n}\cdot\boldsymbol{e_y}
    \label{equ:boundarycondition}
\end{equation}
on the airfoil surface. 
Here, $\boldsymbol{n}$ is the wall normal unit vector, and $\boldsymbol{e_y}$ is the unit vector pointing at the direction of lift. 
By taking the inner product of the momentum equation of the incompressible Navier-Stokes equation with the potential velocity $\boldsymbol{\nabla} \phi_L$, and integrating over the entire fluid domain $V$ and airfoil surface $S$, the lift force can be recovered as
\begin{equation}
    F_L = \int_{V} \boldsymbol{\omega} \times \boldsymbol{u} \cdot \boldsymbol{\nabla}\phi_L \mathrm{d}V 
    + \frac{1}{Re}\int_S\boldsymbol{\omega} \times \boldsymbol{n} \cdot (\boldsymbol{\nabla}\phi_L + \boldsymbol{e_y}) \mathrm{d}S.
    \label{equ:liftElement}
\end{equation}
The integrands in the first and second terms on the right hand side are called the volume ($L_e=\boldsymbol{\omega}\times\boldsymbol{u}\cdot\boldsymbol{\nabla}\phi_L$) and the surface lift elements, respectively. 
At Reynolds number of 1000, the volume force elements contribute more significantly to the total force than the surface force elements.

The vorticity fields $\overline{\omega}_z$ and the volume lift element $L_e$ contours are shown in figures \ref{fig:rL03} and \ref{fig:rC005} for cases with fixed $r_L=0.3$ and fixed $r_C=0.05$, respectively. 
In general, the free shear layers that emanate from the leading and trailing edges of the airfoil generates positive lift, and the flow that attaches on the pressure side generates negative lift.
At fixed $r_L=0.3$, the free shear layer from leading edge (containing negative vorticity) becomes more closer to the suction side of the airfoil as $r_C$ changes from negative to positive, rendering it less separated.
This is reflected in the force element analysis as stronger positive $L_e$ on the suction side.
In addition, the positive vorticity on the pressure side of the airfoil, as well as the corresponding negative lift element, becomes less dense near the aft part of the airfoil, which corresponds to the drastic change of pressure coefficients at the same location, as observed from figure \ref{fig:pressure}($a$).
Due to the augmented positive $L_e$ on the suction side and reduced negative $L_e$ on the pressure side, the airfoil presents considerable lift enhancement with increasing $r_C$.

For an airfoil with fixed trailing camber ratio $r_C=0.05$, not much difference is discerned for the positive lift elements on the suction side as length ratio $r_L$ is varied.
However, for airfoil with large $r_L$, the negative lift elements on the pressure side become coarser towards the aft part, contributing positively to the lift generation.
Essentially, with increasing $r_C$ and $r_L$, the airfoil as a whole becomes more cambered. 
The above analysis shows that the lift enhancement of such cambered airfoil is mostly due to the flow modification near the trailing edge.

\section{Conclusions}
\label{sec:conclusions}
A large number of two-dimensional direct numerical simulations are performed for incompressible flow over a class of hybrid piezocomposite airfoils, at a fixed Reynolds number of 1000.
By varying the length ratio $r_L$ and the camber ratio $r_C$ of trailing control surface, and the incidence angle $\alpha_i$, the hybrid airfoil design results in wide range of angles of attack, encompassing different wake states such as steady flow, periodic shedding, and quasi-periodic shedding.
Compared to a NACA 0012 airfoil, the transition between these wake states can occur that lower angle of attack in the presence of a trailing control surfaces.
The lift coefficient generally increases with $r_C$ and $r_L$ over the studied range of parameters. 
The drag coefficients among cases of different $r_C$ and $r_L$ remain close to each other over a wide range of angle of attack.
By considering cases at a fixed angle of attack of $10^{\circ}$ but with different configurations, the mechanism of lift enhancement with increasing $r_C$ and $r_L$ is explained through pressure distribution and force element method.
These analyses suggest that in the case of increasing $r_C$ at a fixed $r_L$, the flow over both the suction becomes closer to the airfoil surface, providing positive lift. In addition, the flow on the pressure side is also modified in a favorable way that results in higher lift.
On the other hand, with increasing $r_L$ at a fixed $r_C$, only the flow on the pressure side is modified near the trailing edge to enhance lift.
These results provided an improved understanding of low-Reynolds-number aerodynamic characteristics of hybrid airfoils with piezocomposite control surfaces. 
The insight obtained from this research could potentially aid the design of active control methods for small unmanned aerial vehicles.

\section*{Acknowledgments}
The authors are grateful for the computational resources on Amarel cluster provided through the Office of Advanced Research Computing (OARC) of Rutgers University.
This work is partially supported by the Department of Energy (DOE) Advanced Research Projects Agency-Energy (APAR-E) Program award DE-AR0001186.
\bibliography{sample}

\end{document}